\documentclass[twocolumn]{aastex631}

\usepackage{graphicx}
\usepackage{gensymb}
\usepackage{breqn}
\usepackage{amsmath}
\usepackage{soul}

\newcommand*\subtxt[1]{_{\textnormal{#1}}}
\DeclareRobustCommand\_{\ifmmode\expandafter\subtxt\else\textunderscore\fi}

\begin{document}

\title{Traversing the Star-Forming Main Sequence with Molecular Gas Stacks of z$\sim$1.6 Cluster Galaxies}

\author[0000-0001-9369-6921]{Alex Pigarelli}
\affiliation{School of Earth \& Space Exploration, Arizona State University, Tempe, AZ, 85287, USA}
\affiliation{Beus Center for Cosmic Foundations, Arizona State University, Tempe, AZ, 85287, USA}

\author[0000-0003-1832-4137]{Allison Noble}
\affiliation{School of Earth \& Space Exploration, Arizona State University, Tempe, AZ, 85287, USA}
\affiliation{Beus Center for Cosmic Foundations, Arizona State University, Tempe, AZ, 85287, USA}

\author[0000-0001-5851-1856]{Gregory Rudnick}
\affiliation{The University of Kansas, Department of Physics and Astronomy, 1251 Wescoe Hall Drive, Lawrence, KS 66045, USA}

\author[0000-0003-0289-2674]{William Cramer}
\affiliation{School of Earth \& Space Exploration, Arizona State University, Tempe, AZ, 85287, USA}
\affiliation{Department of Physics and Astronomy, Notre Dame University, South Bend, IN 46617, USA}

\author[0000-0002-8909-8782]{Stacey Alberts}
\affiliation{Steward Observatory, University of Arizona, 933 North Cherry Avenue, Tucson, AZ 85719, USA}

\author[0000-0002-3196-5126]{Yannick Bahe}
\affiliation{School of Physics and Astronomy, University of Nottingham, University Park, Nottingham NG7 2RD, UK}
\affiliation{Laboratory of Astrophysics, Ecole Polytechnique F\'{e}d\'{e}rale de Lausanne (EPFL), Observatoire de Sauverny, 1290 Versoix, Switzerland}

\author[0000-0001-9394-6732]{Patrick S. Kamieneski}
\affiliation{School of Earth \& Space Exploration, Arizona State University, Tempe, AZ, 85287, USA}

\author[0009-0004-0000-9539]{Sebastian Monta\~no}
\affiliation{School of Earth \& Space Exploration, Arizona State University, Tempe, AZ, 85287, USA}

\author[0000-0002-9330-9108]{Adam Muzzin}
\affiliation{Department of Physics and Astronomy, York University, 4700 Keele Street, Toronto, Ontario, ON MJ3 1P3, Canada}

\author[0000-0002-7356-0629]{Julie Nantais}
\affiliation{Facultad de Ciencias Exactas, Departamento de F\'isica y Astronom\`ia, Instituto de Astronom\'ia, Fern\'andez Concha 700, Edificio C-1, Piso 3, Las Condes, Santiago, Chile}

\author[0009-0009-9329-1612]{Sarah Saavedra}
\affiliation{School of Earth \& Space Exploration, Arizona State University, Tempe, AZ, 85287, USA}

\author[0000-0002-6327-5154]{Eelco van Kampen}
\affiliation{European Southern Observatory, Karl-Schwarzschild-Str. 2, 85748, Garching bei M\"unchen, Germany}

\author{Tracy Webb}
\affiliation{Department of Physics, McGill Space Institute, McGill University, 3600 rue University, Montr\'eal, Qu\'ebec, Canada, H3A 2T8}

\author[0000-0003-2919-7495]{Christina C. Williams}
\affiliation{NSF National Optical-Infrared Astronomy Research Laboratory, 950 North Cherry Avenue, Tucson, AZ 85719, USA}
\affiliation{Steward Observatory, University of Arizona, 933 North Cherry Avenue, Tucson, AZ 85719, USA}

\author[0000-0002-6572-7089]{Gillian Wilson}
\affiliation{Department of Physics, University of California Merced, 5200 Lake Road, Merced, CA 95343, USA}

\author[0000-0003-4935-2720]{H. K. C. Yee}
\affiliation{The David A. Dunlap Department of Astronomy and Astrophysics, University of Toronto, 50 St George St., Toronto, ON M5S 3H4, Canada}



\begin{abstract}
The cluster environment has been shown to affect the molecular gas content of cluster members, yet a complete understanding of this often subtle effect has been hindered due to a lack of detections over the full parameter space of galaxy star formation rates and stellar masses.  Here we stack CO(2-1) spectra of $z\sim1.6$ cluster galaxies to explore the average molecular gas fractions of galaxies both at lower mass (log($M_{*}/M_{\odot})\sim9.6$) and further below the Star Forming Main Sequence (SFMS; $\Delta$MS$\sim-0.9$) than other literature studies; this translates to a $3\sigma$ gas mass limit of $\sim7\times10^9\,M_{\odot}$ for stacked galaxies below the SFMS. We divide our sample of 54 $z \sim 1.6$ cluster galaxies, derived from the Spitzer Adaptation of the Red-Sequence Cluster Survey, into 9 groupings, for which we recover detections in 8. The average gas content of the full cluster galaxy population is similar to coeval field galaxies matched in stellar mass and star formation rate. However, when further split by CO-undetected and CO-detected, we find that galaxies below the SFMS have statistically different gas fractions from the field scaling relations, spanning deficiencies to enhancements from $2\sigma$ below to $3\sigma$ above the expected field gas fractions, respectively. These differences between $z = 1.6$ cluster and field galaxies below the SFMS are likely due to environmental processes, though further investigation of spatially-resolved properties and more robust field scaling relation calibration in this parameter space are required.

\end{abstract}

 \keywords{Galaxy clusters (584) --- High-redshift galaxy clusters (2007) --- Galactic and extragalactic astronomy (563) --- Radio astronomy (1338) --- Molecular Gas (1073)}

\section{Introduction}
 It has been well-established in the literature that a galaxy's environment has an impact on various galactic properties, including morphology \citep{1980ApJ...236..351D,1997ApJ...490..577D,1984ApJ...281...95P} and star formation rate (SFR) \citep{1984ARA&A..22..185D,2002MNRAS.334..673L,2010ApJ...721..193P}.  In recent years, an abundance of molecular gas studies have also looked at the link between environment and its impact on the gas properties of galaxies. Molecular gas, the raw fuel of star formation, provides a unique perspective on galaxy evolution as it probes future star formation in galaxies and it has been shown to be affected by cluster environments. \par

Historically, carbon monoxide (CO) has been used as a tracer of molecular gas in galaxies due to the difficulty of observing molecular hydrogen \citep{2005ARA&A..43..677S, 2013ARA&A..51..105C}.   In particular, several high-resolution studies at low-redshift have investigated the molecular gas reservoirs in cluster galaxies: the Virgo Environment Traced in CO (VERTICO; \citealt{2021ApJS..257...21B}), the ALMA Fornax Cluster Survey (AlFoCS; \citealt{2019MNRAS.483.2251Z}), and the GAs Stripping Phenomena in galaxies survey (GASP; \citealt{2017ApJ...844...48P}).   These surveys explored the environmental effects on cluster galaxies, with all finding evidence of a significant population of cluster galaxies with debris trails (``jellyfish galaxies''), and/or disturbed morphologies and kinematics, most likely as a result of ram-pressure stripping (RPS; \citealt{2019MNRAS.483.2251Z, Moretti_2020a, 2022ApJ...933...10Z}). \par

Aside from morphological changes to the molecular gas in galaxies, environmental effects have also been observed to alter a galaxy's total gas content\footnote{Details of environmental quenching mechanisms can be found in \cite{2021PASA...38...35C, 2022A&ARv..30....3B, 2022Univ....8..554A}}. Some quenching mechanisms strictly remove gas from galaxies, but RPS has been observed to remove gas while also compressing gas on the leading edge of the galaxy, which can create a favorable environment for molecular gas formation \citep{2012A&A...543A..33V, 2017ApJ...839..114J, 2020ApJ...901...95C, 2021ApJ...921...22C, Moretti_2020a, 2022A&A...658A..44R}. RPS can therefore potentially increase the total molecular gas content of a galaxy so that it can be detected through either gas tails, as in jellyfish galaxies, or possibly through elevated gas content.  One such example of RPS forming as well as removing molecular gas is the jellyfish JW100, where the total integrated molecular gas content of the galaxy and tail exceeds the molecular gas content of similar, undisturbed galaxies \citep{Moretti_2020a}.\par

To detect deviations in integrated molecular gas content, one must compare observations of cluster galaxies to those of coeval field galaxies. Large samples of integrated CO in field galaxies across redshift are thus key to assessing the ubiquity of an environmental impact on molecular gas. Work by \cite{2018ApJ...853..179T} found an empirically-derived ``field scaling relation'' (hereafter T18, see also \citealt{2015ApJ...800...20G}) between a galaxy's redshift, stellar mass, SFR, and molecular gas mass among star-forming field galaxies. T18 used 1,444 measurements of molecular gas at $0<z<4$ from many surveys and found the relationship between the galaxy properties and molecular gas content had a fairly narrow scatter (0.2 dex for individual galaxies, 0.1 dex for averages of galaxies).  Therefore, cluster galaxies can be classified as gas-deficient or gas-rich in relation to field galaxies of similar properties.  However, the T18 relation was mainly created from galaxies with SFRs along and above the Star-Forming Main Sequence (SFMS). As such, the field scaling relation is not calibrated for galaxies below the SFMS and does not predict observed gas masses in quiescent galaxies (see \citealt{2021ApJ...908...54W, 2021Natur.597..485W}). 

For galaxies on and around the SFMS, the field scaling relation can be used to investigate trends of gas content in cluster galaxies over redshift. In \cite{2022Univ....8..554A}, this is depicted in Figure 14 for CO measurements in cluster galaxies from $z\sim0-2$. To summarize, in local clusters between $z=0-0.2$, there is a wide range of gas content which spans $\sim$2 dex below to $\sim$0.5 dex above T18. At slightly higher redshift ($0.2<z<0.6$), integrated molecular gas measurements have similarly found cluster galaxies that are gas-deficient as well as galaxies that are gas-rich \citep{2020A&A...640A..65C, 2016MNRAS.459.3287C, 2009MNRAS.395L..62G, 2011ApJ...730L..19G, 2021A&A...654A..69S}. At high redshift ($z>1$), a majority of cluster galaxies are consistent with or gas-rich compared to the T18 field scaling relation and, unlike low redshift results, a population of significantly gas-deficient star-forming galaxies is notably absent \citep{2012MNRAS.426..258A, 2012ApJ...752...91W, 2017ApJ...842L..21N, 2019ApJ...870...56N, 2018ApJ...856..118H, 2022ApJ...929...35W, 2017ApJ...849...27R}. It is not known whether this result simply reflects limited CO sensitivity and Malmquist Bias, or indicates a reversal of what has been observed in the local galaxy clusters. 

In this paper, we address this problem by probing CO emission to fainter luminosities through stacking analyses with 54 galaxy cluster members in three $z\sim1.6$ galaxy clusters from the Spitzer Adaptation of the Red-Sequence Cluster Survey (SpARCS; \citealt{2009ApJ...698.1943W, 2009ApJ...698.1934M}). These clusters are of particular interest as their members are known to have elevated gas fractions \citep{2019ApJ...870...56N}, high levels of kinematic asymmetries \citep{2023ApJ...944..213C}, and a wide range of UV-optical color gradients \citep{2024ApJ...975..144C} compared to field galaxies. On the other hand, the cluster members have also been observed to have H$\alpha$ emission similar to field galaxies \citep{2020MNRAS.499.3061N} . We divide cluster members in different regions of the $M_{*}-\Delta$MS (offset from SFMS) parameter space into groupings of all members, individually CO-detected members, and individually CO-undetected members to obtain the stacked spectra and average molecular gas masses of the galaxies in these bins. Comparing these gas masses to the T18 field scaling relation, we can assess whether the galaxies are, on average, influenced by environmental effects. The evidence of environmental effects could manifest through increased or decreased gas masses compared to coeval field galaxies. \par

We organize the paper as follows: Section \ref{sec:obs} introduces the three $z\sim$1.6 clusters and their observations, Section \ref{sec:data} presents the data preparation, Section \ref{sec:methods} shows our analysis, Section \ref{sec:discussion} contains a discussion of our results, and Section \ref{sec:conclusion} summarizes our conclusions. Throughout this paper, we adopt a Chabrier IMF and a $\Lambda$CDM cosmology with $H_{0}=70$ km s$^{-1}$ Mpc$^{-1}$, $\Omega_{M} = 0.3$, and $\Omega_{\Lambda}=0.7$.

\section{Observations\label{sec:obs}}

\begin{figure*}[!t]
    \centering
    \includegraphics[width=0.97\textwidth]{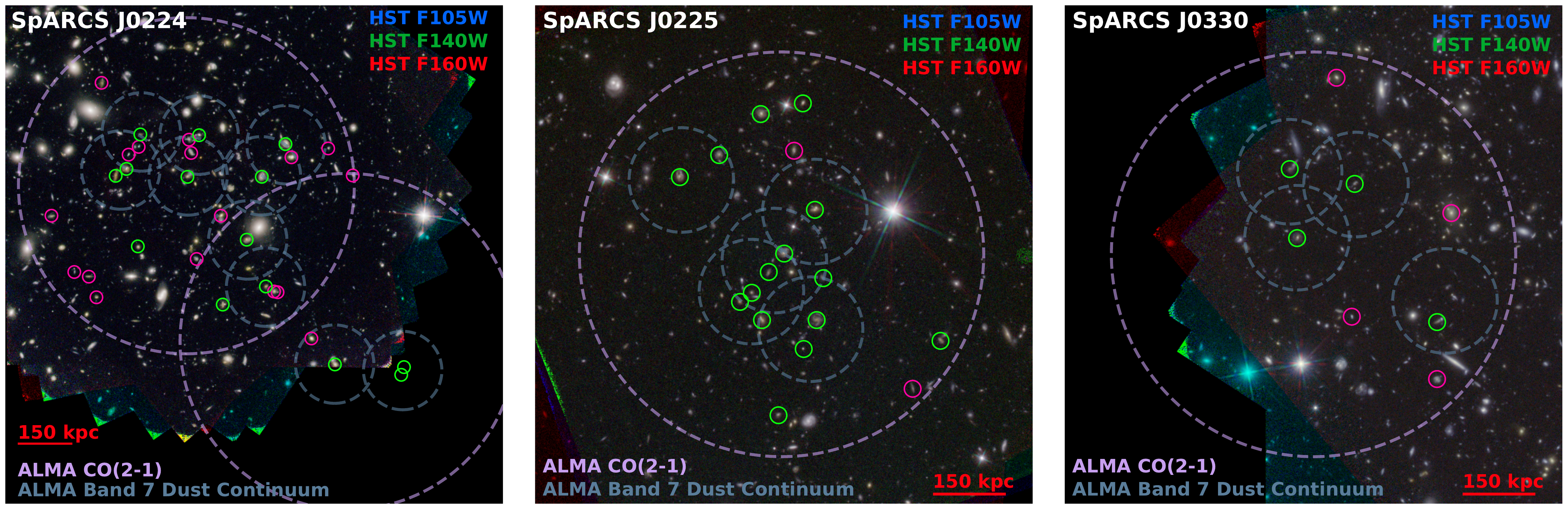}
    \caption{Shown above are the three SpARCS galaxy clusters used in this work. The background RGB images are created from HST F160W, F140W, and F105W imaging. The primary beams of the ALMA Band 3 CO(2-1) imaging and the ALMA Band 7 dust continuum imaging are shown by the purple and blue dashed circles, respectively. The green and pink circles denote the locations of the CO-detected and CO-undetected cluster members, respectively. See Section \ref{subsec:CO_reduction} for further details.}
    \label{fig:cluster_maps}
\end{figure*}

\subsection{SpARCS Clusters\label{subsec:clusters}}
In this paper, we explore the CO content of galaxies within three massive ($M_{*} \gtrsim 10^{14} M_{\odot}$) galaxy clusters at z$\sim$1.6 \citep{2016A&A...592A.161N, 2017ApJ...842L..21N, 2019ApJ...870...56N, 2023ApJ...944..213C}: SpARCS J022416-032330 (``J0224''), SpARCS J022546-035517 (``J0225''), and SpARCS J033056-284300 (``J0330''). These clusters were identified with the Stellar Bump Technique \citep{2013ApJ...767...39M} and have been imaged in 18 photometric bands spanning UV-FIR. All clusters have ground-based ugrizYKs, 3.6/4.5/5.0/8.0/24.0$\mu$m \textit{Spitzer} imaging, and \textit{Hubble Space Telescope} F160W. J0224 and J0330 also benefit from \textit{HST} F105W and F140W imaging from the ``See Change'' program \citep{2021ApJ...912...87H}. Optical and NIR bands were PSF matched to the Ks band images \citep{2016A&A...592A.161N}. In addition, all clusters were a part of the XMM-LSS and CDFS fields imaged by the PACS and SPIRE instruments aboard \textit{Herschel} which have publicly available, deblended FIR catalogs (100/160/250/350/500 $\mu$m). These catalogs were deblended to match the resolution of the $\textit{Spitzer}$ MIPS 24$\mu$m data \citep{2010MNRAS.409...48R,2012MNRAS.424.1614O,2012MNRAS.419..377S,2014MNRAS.444.2870W}. A galaxy is considered a member of one of the clusters if it has a spectroscopic redshift that falls within $\Delta z_{\textrm{cluster}} = \pm 0.015$ from the cluster redshift. Currently, 113 cluster members have been identified over all 3 clusters, primarily from their H$\alpha$ emission \citep{2016A&A...592A.161N}. 

\subsection{ALMA Observations\label{subsec:alma_obs}}
Each cluster was targeted with ALMA Band 3 to capture the CO(2-1) emission within $\pm$3,000 km s$^{-1}$ of the clusters' systematic redshifts. For J0224 and J0330, we use single epoch, spatially-resolved measurement sets (Project code: 2018.1.00974.S, PI: Noble) and for J0225, we combine two spatially-resolved measurement sets (project codes: 2017.1.01228.S, 2021.1.01002.S, PI: Noble). The measurement sets have synthesized beams of approximately 0\farcs5$\times$0\farcs4 ($\sim 4.2\times3.4$ kpc) and total integration times for the clusters are 2.91 and 3.33 hours for two separate pointings in J0224, 11.65 hours combined in a single pointing for J0225, and 3.94 hours in a single pointing for J0330. The imaging contains CO(2-1) detections of 32 cluster members, with an additional 22 spectroscopically-confirmed cluster members not detected, but within the $\sim$95$\arcsec$ diameter primary beams. A cluster member was considered to be detected if it had a CO detection with a 3$\sigma$ peak spanning more than two adjacent 50 km s$^{-1}$ channels with a corresponding Ks-band counterpart. \par

We also use ALMA Band 7 dust continuum observations (project code: 2021.1.01257.S, PI:Noble) which cover 36 of the 54 cluster members present in the CO(2-1) imaging. Each of the 19 pointings spread across the three clusters was targeted at rest frame $\sim$330$\mu$m  and has $\sim$35-40 minutes of integration time with a central rms of 0.02 mJy beam$^{-1}$. The primary beam of these observations is $\sim$25$\arcsec$ and the median synthesized beam size is 0\farcs45$\times$0\farcs37 ($\sim 3.8\times 3.1$ kpc).

Figure \ref{fig:cluster_maps} shows HST imaging of each cluster used in this work along with the primary beams of the ALMA Bands 3 and 7 observations with the CO-detections and CO-non-detections notated.

\section{Data\label{sec:data}}
\begin{figure}[]
    \centering
    \includegraphics[width=0.47\textwidth]{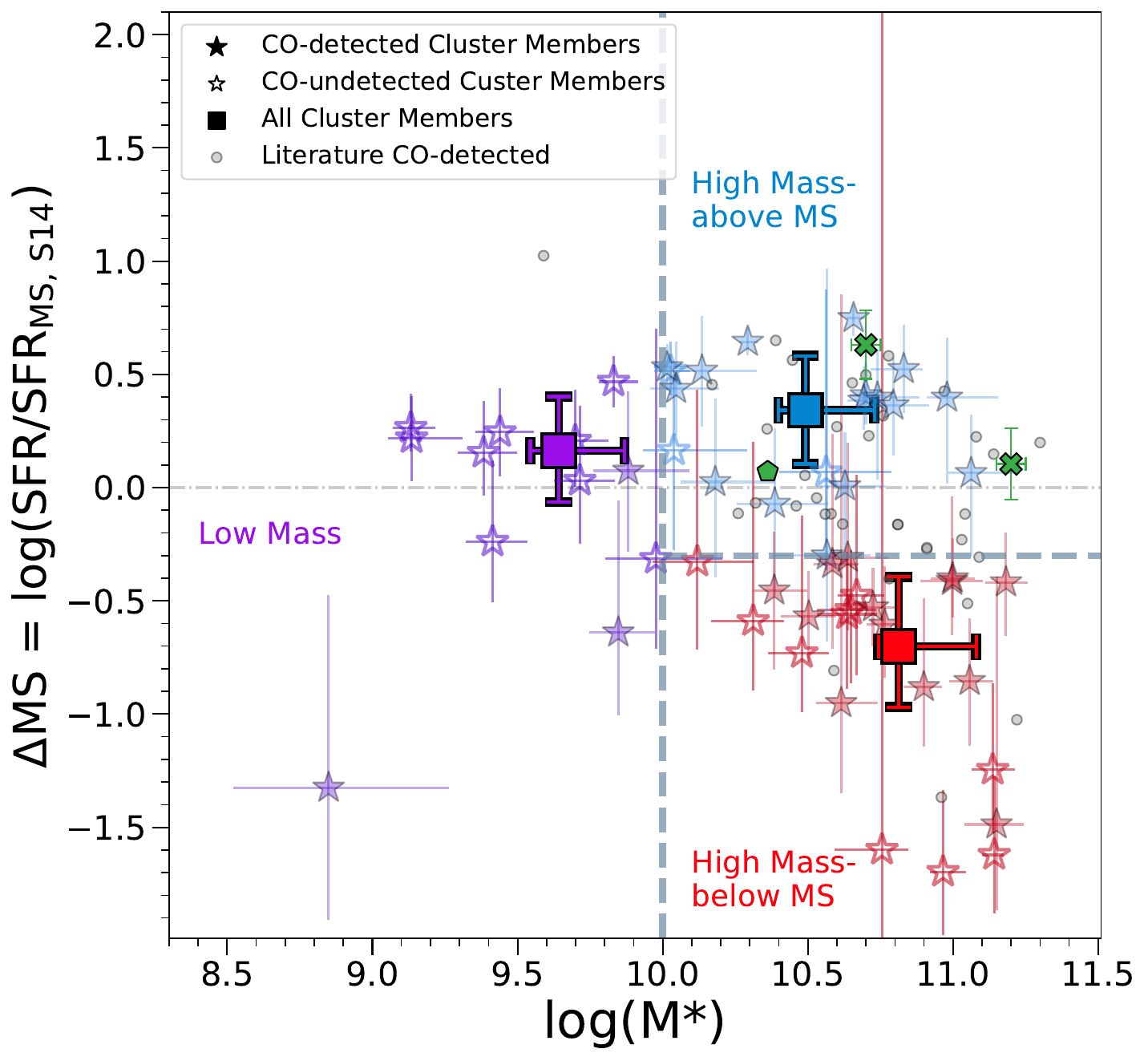}
    \caption{Our sample of cluster galaxies are shown with their stellar masses and offsets from the Star-Forming Main Sequence (SFMS). CO-detected cluster members are shown with filled stars and CO-undetected cluster members with open stars. The averages of cluster members in the three groupings (Low Mass, High Mass-above SFMS, and High Mass-below SFMS) are shown with the solid squares with error bars representing the uncertainty on the mean. Comparison stacks of cluster members from \citealt{2018ApJ...856..118H} (pentagon, $z=1.46$) and \citealt{2022ApJ...927..235A} (crosses, $z\sim1.4$) are shown by the dark green points. Grey circles indicate individual $z>1$ literature cluster members \citep{2022ApJ...929...35W, 2017ApJ...849...27R, 2012MNRAS.426..258A, 2018MNRAS.479..703C, 2018ApJ...856..118H}.}
    \label{fig:sample_binning}
\end{figure}
\subsection{Band 3 CO(2-1) Reduction and Spectral Extraction}\label{subsec:CO_reduction}
We reduce the ALMA measurement sets using the CASA ALMA Science Pipeline version 6.5.4.9 \citep{2007ASPC..376..127M}. For the CO(2-1) observations, the two J0225 measurement sets are first concatenated and then the measurement sets of each cluster are uv-continuum subtracted and cleaned using \texttt{tclean} with natural weighting, a stopping threshold of 2.5$\sigma$, manual masking, primary beam response set to 20\%, and primary beam correction enabled. The synthesized beams for the images are 0\farcs45$\times$0\farcs35 for J0224, 0\farcs52$\times$0\farcs41 for J0225, and 0\farcs48$\times$0\farcs39 for J0330. Each image is created with a cell size chosen such that there are 8 pixels across the minor axis of each of the clean beams. The central rms in 50 km s$^{-1}$ channels are 0.05 mJy beam$^{-1}$, 0.04 mJy beam$^{-1}$, and 0.07 mJy beam$^{-1}$ for J0224, J0225, and J0330, respectively. \par 

We measure the size of each CO-detected cluster member with the \texttt{imfit} task in CASA, which fits a 2D Gaussian profile, on velocity-integrated intensity maps (moment 0 maps) for each galaxy. These sizes and position angles are then used to create custom elliptical spectral extraction apertures for each galaxy at a radius of 4 * Half Width at Half Maximum / 2.355, enclosing the flux within 4$\sigma$. We additionally average the spectra of CO-undetected cluster members shifted to the rest frame to perform a curve of growth analysis and find the total flux from a Gaussian fit flattened beyond a circular aperture radius of 0\farcs8. We use this size aperture for spectral extraction of the CO-undetected galaxies. 

\subsection{Band 7 Dust Continuum Reduction and Flux Extraction}\label{subsec:dust_reduction}
Each ALMA Band 7 pointing, centered at 345 GHz, was individually cleaned with \texttt{tclean} using a natural weighting, a stopping threshold of 2.5$\sigma$, manual masking, primary beam response set to 20\%, and primary beam correction enabled. The minor axes of the clean beams are similar in size which enables a uniform pixel size across the images of 0\farcs037, or $\sim$10 pixels across the minor axes. We use \texttt{imfit} at the locations of known cluster members to measure the integrated flux and require a signal-to-noise of greater than 3 to consider a cluster member detected. For undetected cluster members, we extract blank apertures at the same primary beam response as the cluster member to estimate the 3$\sigma$ upper limit on the emission. While the Band 7 imaging has uniform depth in all pointings, the $3\sigma$ upper limits change per galaxy due to their location in the primary beams (e.g., sources closer to the edge have elevated $3\sigma$ limits as the primary beam correction increases the local pixel-to-pixel rms). \par

\subsection{Stellar Masses and Star Formation Rates}\label{subsec:sm_and_sfr}
We use the SED fitting code \texttt{CIGALE} (Code Investigating GALaxy Emission; \citealt{2019AA...622A.103B}) to estimate stellar masses and SFRs of the SpARCS cluster members with all available bands of photometry. We chose to use \texttt{CIGALE} for its optical-IR energy balance and ability to consider upper limits when sources were undetected in an imaging band. Due to the shallow HerMES imaging depth, when a source had no FIR emission in the catalog, an upper limit of the 3$\sigma$ clipped rms noise was used for each band as described in \cite{2010MNRAS.409...48R}. For the population of SpARCS cluster members greater than 0.3 dex below the SFMS (3 galaxies have log($M_{*}/M_{\odot}) <$ 10.0, 23 have log($M_{*}/M_{\odot}) > 10.0$), the SFRs are constrained by having HerMES upper limits on their FIR emission. Of these galaxies, 8 of 26 have ALMA Band 7 dust continuum detections with an average uncertainty on the measurements of 0.16 mJy and an average detection significance of 7.6$\sigma$. 7 of the 26 have ALMA Band 7 3$\sigma$ upper limits with an average limit of 0.30 mJy.  The remaining 11 galaxies were not contained in the ALMA Band 7 pointings and they only have their FIR emission constrained by the HerMES 3$\sigma$ limits of 10.1 mJy, 10.8 mJy, and 14.5 mJy for the 250$\mu$m, 350$\mu$m, and 500$\mu$m bands, respectively.\par

We assume an exponentially declining star formation history with BC03 simple stellar populations \citep{2003MNRAS.344.1000B}, a Chabrier initial mass function \citep{2003PASP..115..763C}, and Calzetti dust attenuation \citep{2000ApJ...533..682C}. The range of stellar masses from \texttt{CIGALE} spans log$(M_{*}) = 8.8 - 11.2$ and star formation rates span $0.1 - 279$ $M_{\odot}$ yr$^{-1}$ averaged over the last 100 Myr. In Figure \ref{fig:sample_binning}, we show the distribution of stellar masses and offsets from the SFMS (\citealt{2014ApJS..214...15S}, hereafter S14) of our galaxy sample as well as our grouping criteria of galaxy properties (presented in Section \ref{subsec:grouping}). In this parameter space, we also show $z > 1$ literature CO-detected galaxies and two other z$\sim$1.5 literature stacking studies \citep{2018ApJ...856..118H, 2022ApJ...927..235A} to illustrate that our work effectively probes new regions of $M_{*}$ - $\Delta$MS parameter space that other high-redshift cluster studies have not. We note that a vast majority of literature CO-detected galaxies at $z>1$ (28 of 35) fall into the upper right portion of this parameter space (see Figure \ref{fig:sample_binning}) which includes high-mass star-forming galaxies. Meanwhile, 35 of our 54 SpARCS cluster members ($\sim$65\%) are outside of this high-mass star-forming classification.

\begin{center}
\begin{table*}[!t]
    \centering
    \begin{tabular}{l|c | c c c c c | c}
        \textbf{Stack} &  $N_{gal}$ & \textbf{Stellar Mass}  & \textbf{SFR}  & \textbf{Integrated Flux}\footnote{The integrated flux under a Gaussian fit to the stacked spectrum of galaxies. The uncertainties reported correspond to the $1\sigma$ bootstrapped variance of realizations of the grouping; see Section \ref{subsec:bootstrapping}} & \textbf{Gas Mass} & \textbf{$f_{gas}$}\footnote{$f_{gas}= M_{gas} / (M_{gas} + M_{*})$} & $\Delta$T18\footnote{$\Delta$T18 = log($\mu_{measured}$) - log($\mu_{T18}$); $\mu = M_{gas} / M_{*}$} \\
         &  & ($\times$10$^{10}$ M$_{\odot}$) & (M$_{\odot}$ yr $^{-1}$) & (Jy km s$^{-1}$) & ($\times$10$^{10}$ M$_{\odot}$) & & \\\hline \hline
        LM - all & 12 &  0.4$_{-0.1}^{+0.1}$&  13$^{+4}_{-3}$ & 0.04$^{+0.03}_{-0.02}$& 1.2$^{+0.9}_{-0.6}$& 73$^{+15}_{-13}$\% & 0.09\\
        LM - CO & 3 &  0.5$_{-0.1}^{+0.2}$& 7$_{-4}^{+4}$ & 0.15$_{-0.01}^{+0.01}$& 4.3$^{+0.3}_{-0.3}$ & 90$^{+3}_{-4}$\% & 0.82\\
        LM - no CO & 9 & 0.4$_{-0.1}^{+0.1}$& 16$_{-3}^{+4}$ &  $<$0.03 & $<$0.9& $<$67\%& $<$-0.24\\ \hline
        HM-aMS - all & 19 & 3.1$_{-0.7}^{+0.9}$& 78$_{-16}^{+17}$ & 0.30$^{+0.06}_{-0.06}$& 6.3$^{+1.3}_{-1.1}$ & 67$^{+6}_{-8}$\% & 0.13\\
        HM-aMS - CO & 16 & 3.1$_{-0.6}^{+0.8}$& 82$_{-16}^{+11}$ & 0.31$_{-0.06}^{+0.08}$& 6.8$^{+1.7}_{-1.3}$ & 69$_{-7}^{+7}$\% & 0.14\\
        HM-aMS - no CO & 3 & 3.3$_{-1.2}^{+1.6}$& 47$_{-19}^{+67}$ & 0.10$^{+0.01}_{-0.03}$& 2.0$^{+0.2}_{-0.6}$& 39$_{-14}^{+9}$\% & -0.24\\ \hline
        HM-bMS - all & 23 & 6.5$_{-1.0}^{+1.2}$& 15$_{-6}^{+8}$ & 0.17$^{+0.06}_{-0.04}$& 3.3$^{+1.2}_{-0.8}$ & 34$_{-7}^{+9}$\% & 0.17\\
        HM-bMS - CO & 13 & 6.2$_{-1.1}^{+1.2}$ & 15$_{-6}^{+7}$ & 0.30$^{+0.08}_{-0.04}$ & 5.9$^{+1.9}_{-0.8}$ & 49$_{-6}^{+8}$\% & 0.40\\
        HM-bMS - no CO & 10 & 7.0$_{-1.0}^{+1.2}$& 12$_{-6}^{+8}$ & 0.05$^{+0.02}_{-0.01}$ & 1.0$^{+0.4}_{-0.2}$ & 12$_{-3}^{+5}$\% & -0.32\end{tabular}
    \caption{A summary of the weighted average properties from CIGALE SED fitting and spectral measurements.}
    \label{tab:SED_properties}
\end{table*}
\end{center}

\section{Analysis} \label{sec:methods}
\subsection{Spectral Stacking}\label{subsec:grouping}
With the extracted spectra from the CO(2-1) image cubes as described in Section \ref{subsec:CO_reduction}, we shift the spectra to the rest frame based on their CO redshift, when available, or H${\alpha}$ spectroscopic redshift \citep{2016A&A...592A.161N}. We then interpolate the spectra to common 50 km s$^{-1}$ velocity bins, average the interpolated spectra, and perform an initial $\chi^{2}$ Gaussian fit to generate a $\pm5\sigma$ bound on the emission line velocity center. A line-free channel-to-channel rms value is calculated for each spectrum outside of the $\pm 5 \sigma$ bounds which provides the weightings for each spectrum when they are stacked with inverse-variance weighting. We apply these same weightings when calculating the weighted average stellar masses and weighted average SFRs.\par

We then separate the spectra into three groupings based on stellar mass and offset from the SFMS and stack the spectra in each grouping with inverse-variance weighting according to their line-free channel-to-channel rms. S14 reports the scatter on a majority of literature SFMS fits to be between 0.2-0.3 dex (T18 reports this conservatively as 0.3 dex). To preserve a pristine sample of galaxies below the SFMS, we adopt a scatter of 0.3 dex on the SFMS and classify cluster members greater than 0.3 dex below the Main Sequence to be ``below Main Sequence'' galaxies. For our first classification, all galaxies with a stellar mass of log($M_{*}/M_{\odot}) < 10.0$ are assigned to the Low Mass group (LM grouping). Galaxies with stellar masses of log($M_{*}/M_{\odot}) > 10.0$ that are above or consistent with the SFMS ($\Delta$MS $>$ -0.3) are considered High Mass above-SFMS galaxies (HM-aMS grouping). While not all galaxies in this grouping are actually above the SFMS, only 3 are below, and 13 of 19 galaxies have offsets greater than +0.3 dex (see Figure \ref{fig:sample_binning}). The remaining galaxies that are above log($M_{*}/M_{\odot}) > 10.0$ and more than 0.3 dex below the SFMS are considered part of the High Mass-below SFMS grouping (or HM-bMS). We note that our results are insensitive to adjusting the below Main Sequence classification to any measurement below the Main Sequence ($<$0 dex offset) and choose to keep our classification limit at $<$-0.3 dex offset for a more pristine sample of below Main Sequence galaxies. Both the LM and HM-bMS classifications described here represent relatively unstudied high-redshift cluster populations (see Figure \ref{fig:sample_binning} with 1 literature cluster member in the LM grouping and 6 in the HM-bMS grouping) and therefore provide new insight into the molecular gas masses of cluster members.\par 

The above classifications assign 3 CO-detected and 9 CO-undetected galaxies to the Low Mass grouping, 16 CO-detected and 3 CO-undetected galaxies to the High Mass above-SFMS grouping, and, remarkably, 13 CO-detected and 10 CO-undetected galaxies to the High Mass-below SFMS grouping. We create inverse-variance weighted stacked spectra for each of the 3 classifications of galaxies as well as inverse-variance weighted stacks of spectra for the CO-detected and CO-undetected galaxies within each of the groupings for comparison.

\subsection{Spectral Bootstrapping}\label{subsec:bootstrapping}
We account for the variance of our sample of cluster galaxies and photometric uncertainty by bootstrapping our inverse-variance weighted stacked spectra. In our bootstrapping, we first randomly select the same number of spectra as there are in each grouping, allowing for repetition, and apply inverse-variance weighting to the constituent spectra to create a new ``bootstrapped'' spectrum. This creates a single new realization of our data. We do not allow for re-mixing of the spectra, such that spectra from the low mass grouping are not selected in any of the other groupings and vice versa. This bootstrapping is performed with 1,000 realizations per grouping and the standard deviation in each channel across all realizations gives the uncertainty of the flux for each channel (e.g. the standard deviation of the 1,000 realizations in the -100 km/s channel gives the 1$\sigma$ uncertainty on the -100 km/s channel). We use these uncertainties with the non-bootstrapped stacked spectra and each bootstrapped realization for integrated flux measurements with a Bayesian fitter (see Section \ref{subsec:bayesian}).\par

We note that the uncertainty from the bootstrapping does not yield the final uncertainty on our measured gas masses but constitutes the spectral uncertainty for Bayesian fitting as described in Sections \ref{subsec:bayesian} and \ref{subsec:gas_mass}. Additionally, the number of bootstraps performed does not affect the significance of our final detections as long as there is sufficient sampling of the constituent spectra in each grouping. In line-free channels, the uncertainty on the flux measurements from this methodology accounts for the photometric uncertainty, essentially the equivalent of stacking blank apertures. In channels where there is emission, the bootstrapping simultaneously accounts for both photometric uncertainty and variance of CO flux from the galaxies.\par

We assume that any channel-to-channel correlated noise is minimal and does not impact the spectra. The native velocity resolutions of the ALMA observations are $\sim$13 km s$^{-1}$ which gives $\sim$4 native channels mapped to each 50 km s$^{-1}$ final channel. We bin our final images at 50 km s$^{-1}$ to increase the signal-to-noise ratio in each channel while still having a sufficient number of velocity channels for adequate Gaussian emission line fitting and to be consistent with previously published work on these measurement sets (see \citealt{2019ApJ...870...56N, 2023ApJ...944..213C}). While the native velocity resolutions are not perfect integer multiples of the final velocity resolution, we assume that the large number of native channels mapped to final channels is sufficient to not need to consider channel-to-channel correlated noise (see further discussion of spectral regridding and rebinning in \citealt{2021ApJS..255...19L}).\par

\subsection{Bayesian Fitting of Stacked Spectra\label{subsec:bayesian}}
\begin{figure*}[]
    \centering
    \includegraphics[width=0.95\linewidth]{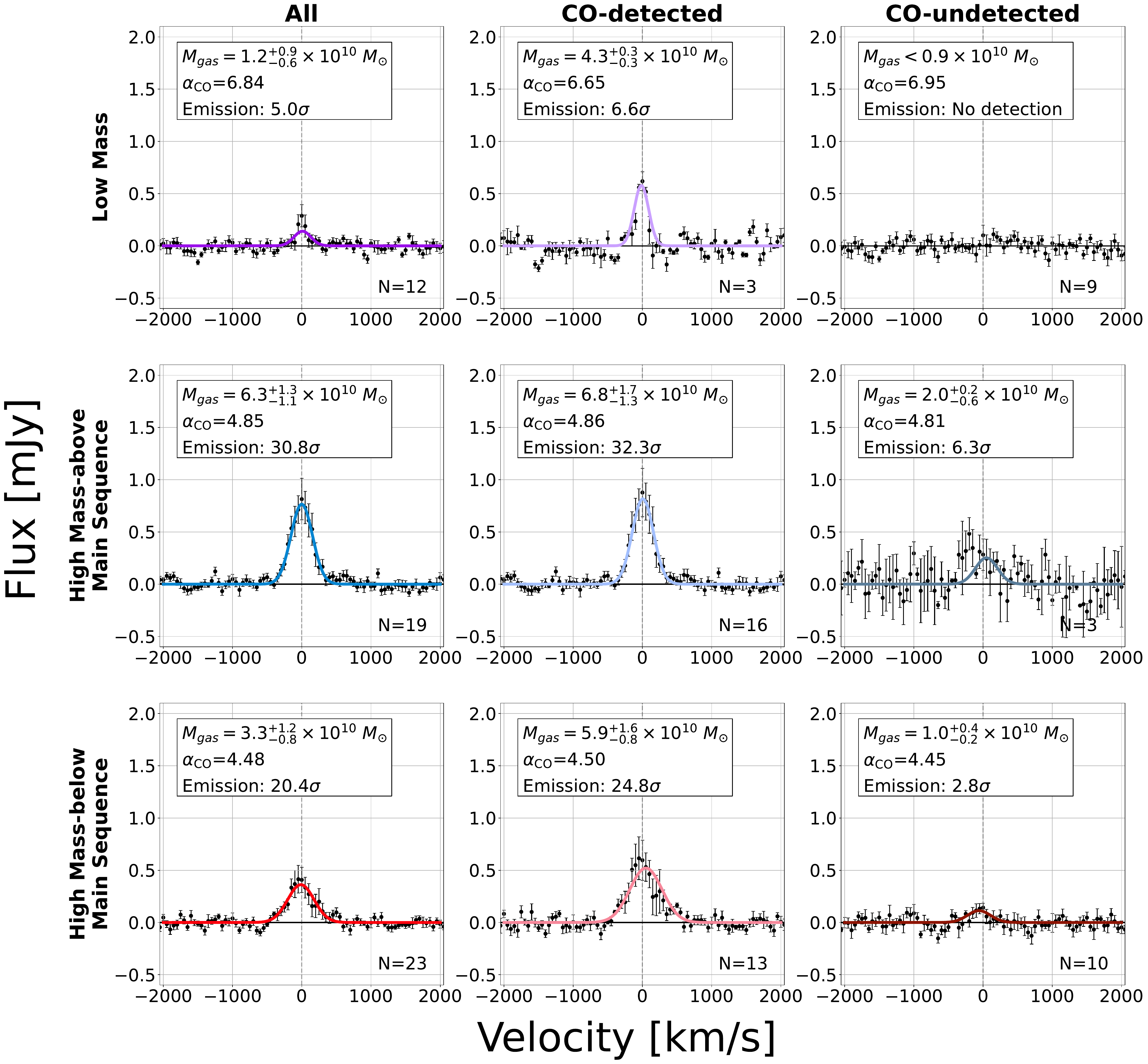}
    \caption{The inverse-variance weighted stacked spectra of LM (top row), HM-aMS (middle row), and HM-bMS (bottom row) groupings. In each case, the left-hand panel shows the full stack, while the middle and right-hand panels only include CO-detected and CO-undetected galaxies, respectively.  In each panel (except the low mass CO-undetected grouping in the top-right corner), the best-fit Gaussian is shown by the solid curve. Stellar masses, SFRs, integrated fluxes, and gas fractions of each grouping are shown in Table \ref{tab:SED_properties}. We report the non-parametric significance of emission in each spectrum as well as the measured gas mass and variance from bootstrapped realizations in each grouping; see Section \ref{subsec:bayesian} for details.}
    \label{fig:stacked_spectra}
\end{figure*}

We model a Gaussian emission line using the \texttt{dynesty} package \citep{2020MNRAS.493.3132S} to perform nested sampling fits to the observed stacked spectra (described in Section \ref{subsec:bootstrapping}) shown in Figure \ref{fig:stacked_spectra}. The prior bounds on our Gaussian emission line fit are: velocity centroid $\in$ (-300,300) km s$^{-1}$; centroid amplitude $\in$ (0.00, 1.50) mJy; FWHM $\in$ (0.01,500) km s$^{-1}$. We assign a uniform prior to the amplitude parameter, and Gaussian priors to the centroid and FWHM parameters. For the centroid prior, we set $\sigma=63.7$ km s$^{-1}$ centered at 0 km s$^{-1}$. We perform initial $\chi^{2}$ fits to the three ``all'' groupings to inform the FWHM priors. For the LM grouping, we set the Gaussian prior on the FWHM to be centered on 300 km s$^{-1}$ with $\sigma=$31 km s$^{-1}$. We give the HM-aMS grouping a Gaussian FWHM prior centered on 365 km s$^{-1}$ with a width of 21 km s$^{-1}$, and the HM-bMS a Gaussian FWHM prior centered on 407 km s$^{-1}$ with a width of 61 km s$^{-1}$.\par

We use this Bayesian fitter to also measure the uncertainty on the flux due to population variance. To do this, we fit the 1,000 bootstrap realizations in each grouping along with the bootstrapped channel-to-channel $1\sigma$ uncertainties (see Section \ref{subsec:bootstrapping}) with the Bayesian nested sampler to obtain the posterior distribution of areas under the Gaussian curves. This quantifies the range of possible fluxes present in each grouping. Second, with the $1\sigma$ bootstrapped uncertainties, we use the observed (non-bootstrapped) weighted spectra, with the same channel-to-channel $1\sigma$ bootstrapping uncertainties, with the Bayesian fitter to measure the average flux of the galaxies in each grouping. In this way, we report a flux measured from the observed stacked spectrum in each grouping, as well as the $1\sigma$ range of bootstrapped fluxes each measured independently with a Gaussian. This range of bootstrapped fluxes is not equivalent to a detection significance, but flux uncertainty due to population variance. As such, we also report a non-parametric emission significance for each observed stacked spectrum which uses the number of emission channels and the rms in line-free channels to assess the significance of emission as shown below \citep{2022MNRAS.514.4205S}
\begin{equation}
    SNR = \frac{\sum_{i}^{N_{ch}}S_{i}\Delta v}{\sigma_{rms}\Delta v \sqrt{N_{ch}}}
\end{equation}
where the $N_{ch}$ is the number of channels with emission, $S_{i}$ is the flux in each channel, $\Delta v$ is the velocity width of each channel, and $\sigma_{rms}$ is the line-free channel-to-channel rms. We show the non-parametric emission significance for each grouping in Figure \ref{fig:stacked_spectra}.

\subsection{Gas Masses\label{subsec:gas_mass}}
We follow the \cite{2005ARA&A..43..677S} method of calculating the CO line luminosity:
\begin{equation}\label{eqn:Lco}
    L'_{\textrm{CO}} = 3.25 \times 10^{7} * S_{\textrm{CO}}\Delta_{\nu} * \frac{D_{L}^{2}}{\nu_{\textrm{rest}}^{2}(1+z)} [\textrm{K km s}^{-1}\textrm{ pc}^{-2}]
\end{equation}
where $S_{CO}\Delta_{\nu}$ is the velocity-integrated flux in Jy km s$^{-1}$, $D_{L}$ is the luminosity distance in Mpc, and $\nu_{rest}$ is the rest frequency of 230.54 GHz. The line luminosity can be converted into a gas mass through:
\begin{equation}\label{eqn:Mgas}
    \frac{M_{gas}}{M_{\odot}} = \alpha_{CO}\frac{L'_{CO}}{r_{21}}.
\end{equation}
Here, we adopt an excitation factor correction, $r_{21}$, of 0.77 \citep{2015A&A...577A..46D,2015ApJ...800...20G}.  The value of the CO-to-H$_{2}$ conversion factor, $\alpha_{CO}$, is difficult to measure outside the Milky Way, and some studies have found it to be metallicity dependent \citep{2014A&A...564A.121C}. \cite{2018ApJ...853..179T} uses a metallicity correction for $\alpha_{CO}$ based on either the observed 12+log(O/H) or a prescription to obtain metallicity from stellar mass and redshift. We do not have metallicity estimates for all CO-detected galaxies in our clusters, but those with measured metallicities fall roughly in range of 12+log(O/H) between $\sim$ 8.4 - 8.7 as estimated from [NII]/H$\alpha$ ratios (Nantais et al. in prep). Therefore, we rely on a stellar mass - metallicity correction to $\alpha_{\textrm{CO}}$ for all of our cluster members. We find the weighted average $\alpha_{CO}$ for our stacks of spectra to be between 4.4 and 6.9 [(M$_{\odot}$ km s$^{-1}$ pc$^{-2}$)$^{-1}$] (see Figure \ref{fig:stacked_spectra}.)\par

With the output from the Bayesian parameter fitting and the assumptions listed above, we detect average gas masses of 1.3$\pm$0.4 $\times$10$^{10}$ M$_{\odot}$,  5.6$\pm$0.6 $\times$10$^{10}$ M$_{\odot}$, and 3.1$\pm$0.4 $\times$10$^{10}$ M$_{\odot}$ for the LM, HM-aMS, and HM-bMS groupings. These gas masses correspond to gas fractions ($f_{gas} = M_{gas}/(M_{gas}+M_{*})$) of 75$\pm$11\%, 65$\pm$8\%, and 32$\pm$6\% respectively. Average properties and gas properties of all groupings of galaxies are presented in Table \ref{tab:SED_properties}.  We compute the offset from the T18 field scaling relation ($\Delta$T18) as the log gas-to-stellar ratio offset ($\mu = M_{gas}/M_{*}$; $\Delta$T18 = log($\mu_{\textrm{measured}}$) - log($\mu_{\textrm{T18}}$)) for each grouping of galaxies given their redshift, stellar mass, and SFRs and also present them in Table \ref{tab:SED_properties}.

\subsection{Gas Detection Limits\label{subsec:upper_limits}}
We investigate the upper limit of gas content that we are able to detect by simulating a flat spectrum, adding a known amount of flux to the spectrum in the form of a Gaussian emission line, adding noise to the data, adding error bars to each channel, and finally attempting to recover the injected flux through the same Bayesian parameter estimation as Section \ref{subsec:bayesian}. We run a full grid of parameters where we vary the channel-to-channel noise of the spectrum, the magnitude of the error bars, and the known amplitude and FWHM of the simulated emission line. \par

\begin{figure}[]
    \centering
    \includegraphics[width=0.45\textwidth]{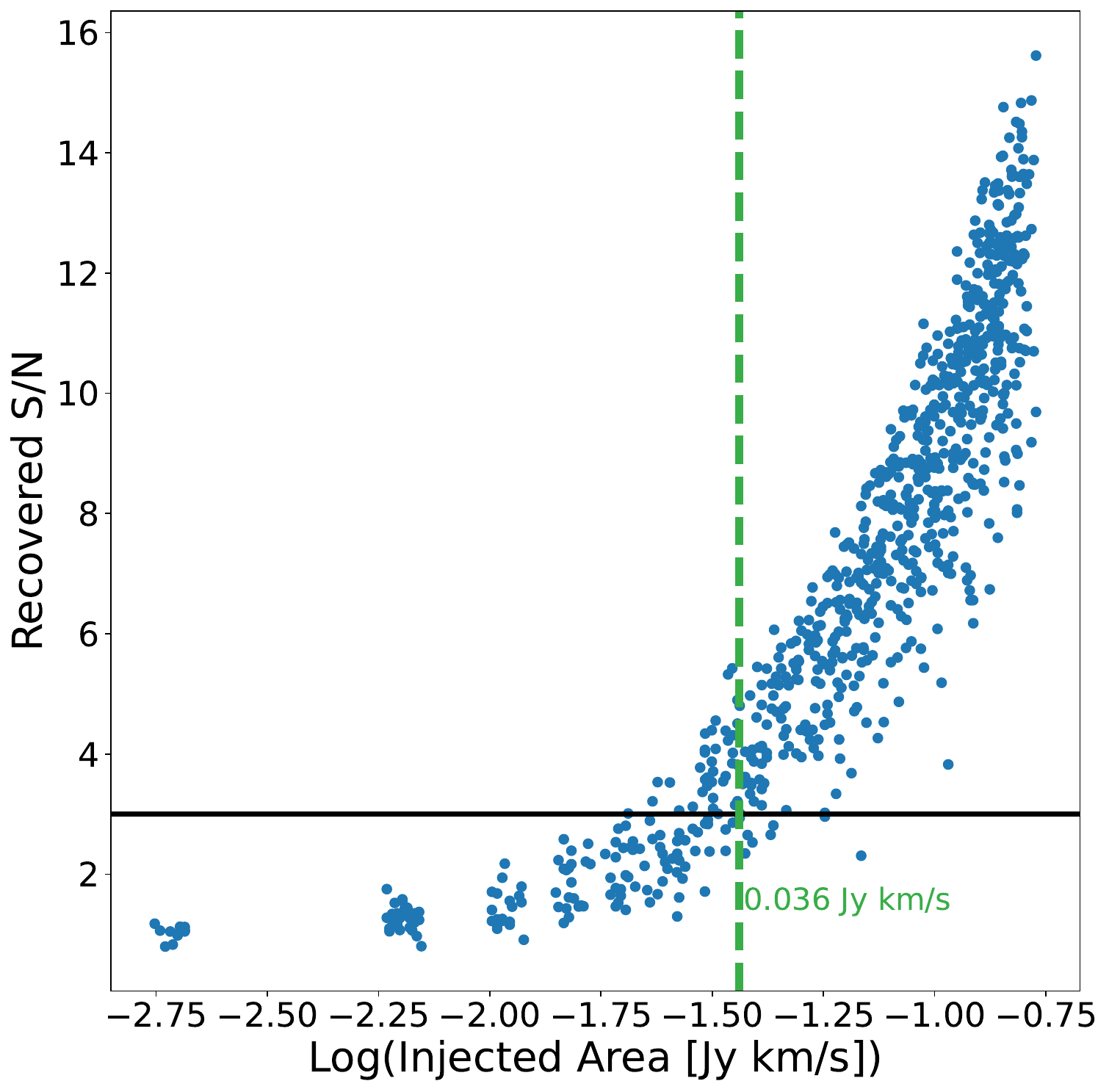}
    \caption{The scatter of recovered S/N as a function of flux injected into a flat spectrum with spectral noise properties similar to our CO-undetected HM-aMS grouping. We find our $3\sigma$ completeness limit to be 0.036 Jy km s$^{-1}$ for this grouping using the procedure described in Section \ref{subsec:upper_limits}.}
    \label{fig:BR}
\end{figure}

To measure the smallest integrated flux we are sensitive to given the noise properties of our spectra, we first filter the grid of parameters to match the properties of our stacked spectra. For each grouping, we select a subset of all the simulated spectra with channel-to-channel noise and error bar magnitude input grid points similar to that of the measured quantities of the weighted spectra. We then compare the signal-to-noise of the Bayesian recovered flux to the injected flux and find the lower limit on the injected flux where 90\% of the stacked spectra have a 3$\sigma$ detection. We consider this flux to be our 3$\sigma$ completeness limit. Figure \ref{fig:BR} shows the signal-to-noise of the recovered flux as a function of the injected flux from the subset of simulated spectra with noise properties similar to the HM-aMS CO-undetected grouping. Starting on the right side of the plot, we take a vertical slice with a width of 2 mJy km s$^{-1}$ and check the percentage of simulated spectra in that slice with signal-to-noise greater than 3. If more than 90\% of simulated spectra in the slice have a signal-to-noise on the flux measurement greater than 3, we move one slice to the left and repeat until a slice no longer has 90\% of the simulated spectra above a signal-to-noise ratio of 3. For the HM-aMS CO-undetected grouping, this completeness limit occurs at 36 mJy km s$^{-1}$.\par

For our ``all'' groupings of galaxies, we measure sensitivity limits of +0.04, -0.76, and -0.46 dex relative to the T18 relation for the LM, HM-aMS, HM-bMS groupings of galaxies at their weighted average stellar masses and $\Delta$MS. Similarly, the CO-detected groupings of galaxies have $\Delta$T18 limits of +0.05, -0.77, and -0.49 dex. And finally, the CO-undetected groupings have limits of -0.04, -0.70, and -0.41 dex. These detection limits, when using the assumptions in Section \ref{subsec:gas_mass}, correspond to $\sim0.9-1\times10^{10}$ M$_{\odot}$ for the LM groupings, $\sim0.7-0.8\times10^{10}$ M$_{\odot}$ for the HM-aMS groupings, and $\sim 0.7\times10^{10}$ M$_{\odot}$ for the HM-bMS groupings.\par

\begin{figure*}[!h]
    \centering
    \includegraphics[width=0.97\linewidth]{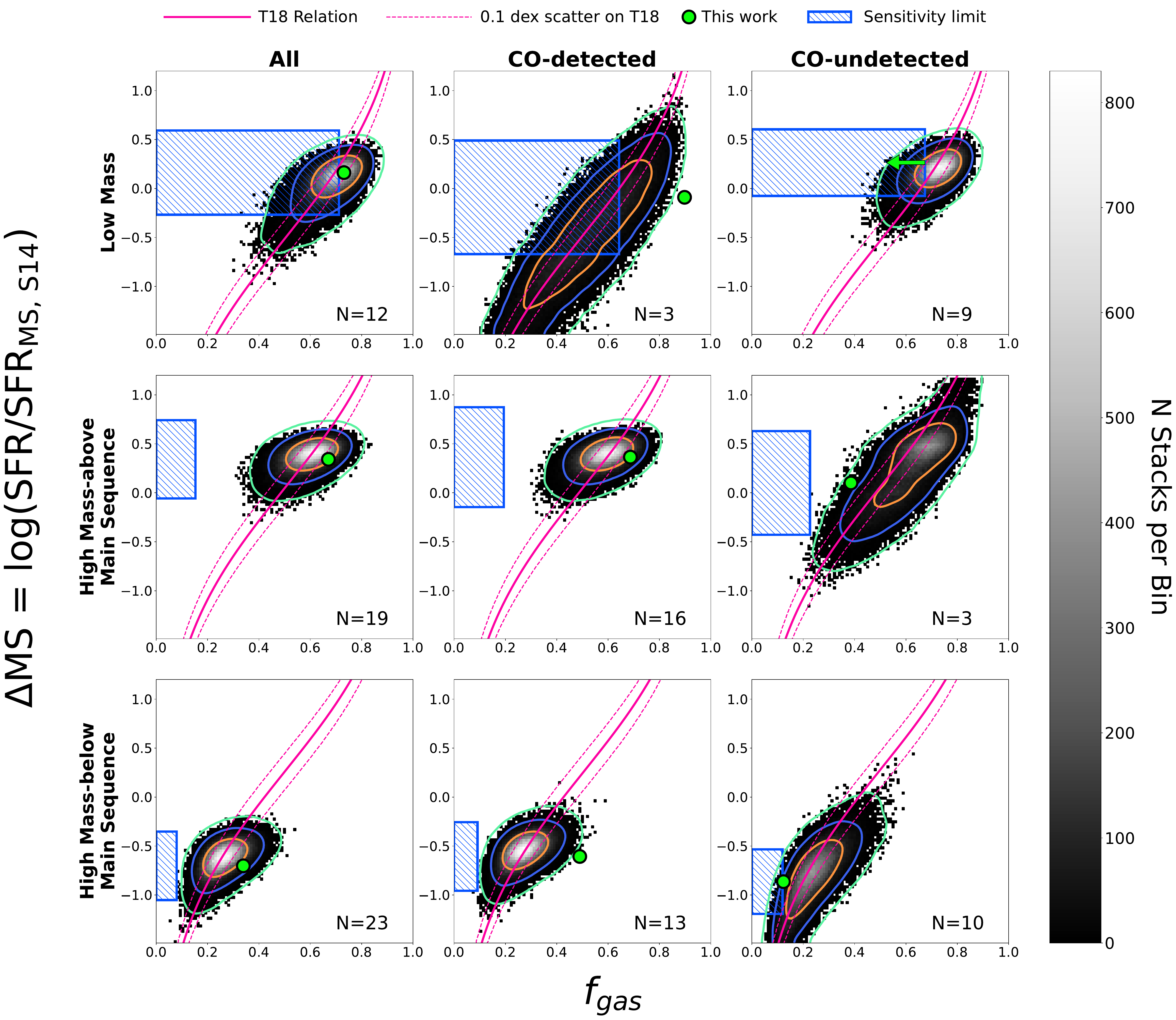}
    \caption{A comparison of our gas fraction measurements to forward-modeled stacks of coeval field galaxies. The 2D histograms in the background of the above plots show the distribution of gas fractions of forward modeled ``field galaxy'' groupings as described in Section \ref{subsec:Forward_modeling}. The variance amongst constituent group members and each stellar mass and/or $\Delta$MS measurements dictates the spread of these histograms in both x and y directions. The green point in each plot shows the measured SFMS offset ($\Delta$MS) and gas fraction of each of our groupings, and the solid pink curve shows the mean field scaling relation for each of our groupings (for the average mass and SFR). The dashed pink lines represent the inherent 0.1 dex scatter on log($M_{gas}/M_{*}$) of the field scaling relations. We show the contours that encompass 68\% (1$\sigma$, orange), 95\% (2$\sigma$, purple), and 99\% (3$\sigma$, green) of groupings of field galaxies. The blue hatched region in each plot shows our sensitivity limit in $f_{gas}$ and the vertical extent depicts the error on the mean $\Delta$MS for each grouping}.
    \label{fig:FM}
\end{figure*}

\begin{figure}
    \centering
    \includegraphics[width=0.45\textwidth]{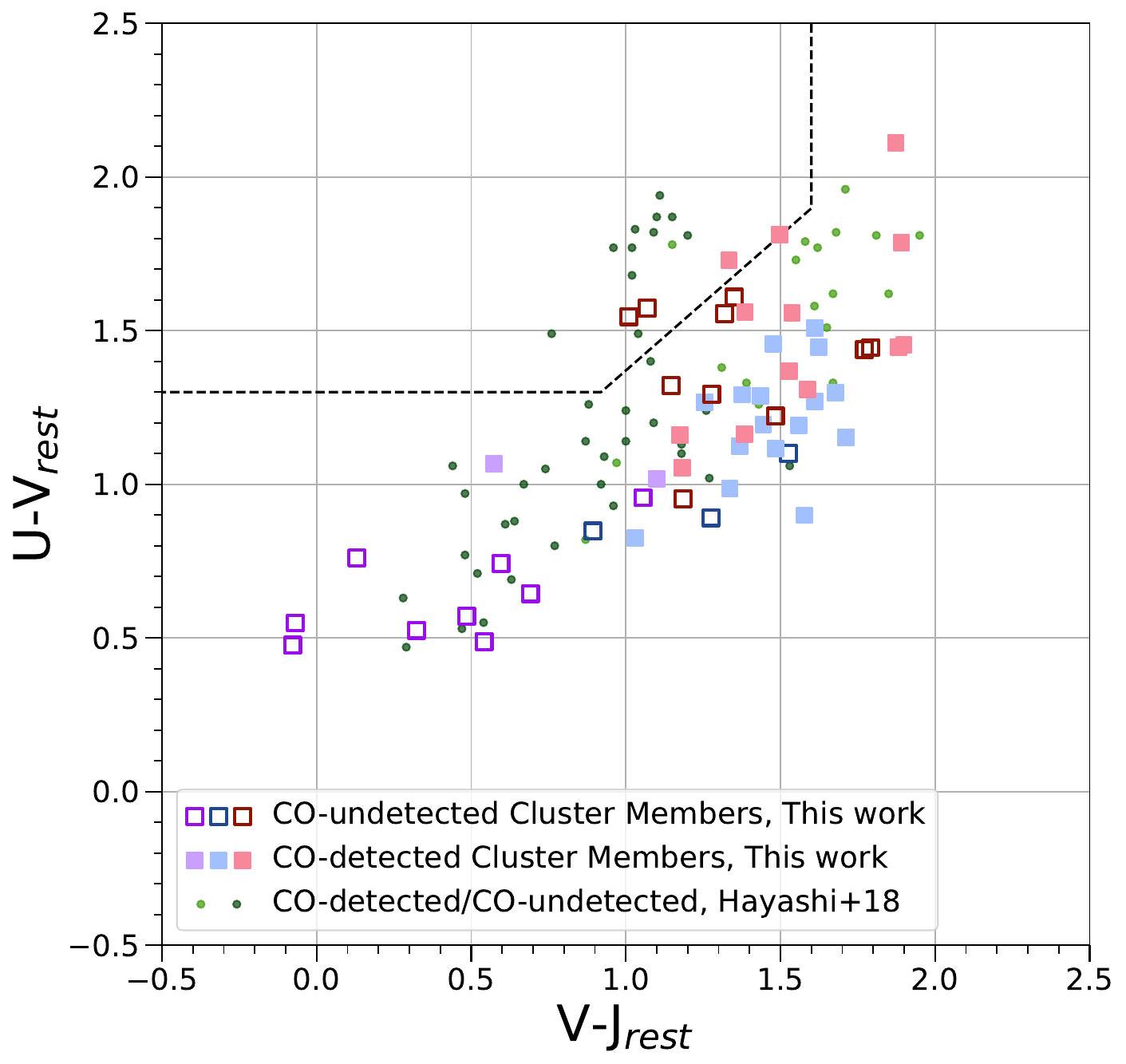}
    \caption{A UVJ diagram of all cluster members. LM galaxies are shown in purple, HM-aMS galaxies in blue, and HM-bMS galaxies in red. Individually CO-undetected and CO-detected galaxies are shown with open and filled points, respectively. Comparable CO-detected and CO-undetected $z=1.46$ cluster members from \cite{2018ApJ...856..118H} are shown by the green points and have a mostly similar distribution in UVJ space to the galaxies studied in this work.}
    \label{fig:uvj}
\end{figure}

\subsection{Forward Modeling\label{subsec:Forward_modeling}}
To determine whether our measured gas fractions are consistent with coeval field galaxies, we employ a forward modeling technique to account for the distribution of galaxy properties within each grouping. We accomplish this by modeling the stellar masses and SFMS offsets of our cluster galaxies, drawing new realizations of the galaxy properties, and inputting them into the T18 field scaling relation to mimic field-like gas fractions.\par
For each grouping, we take the measured stellar masses and SFMS offsets ($\Delta$MS) of each individual galaxy in the grouping and draw a new value within the asymmetric errors 10,000 times. Just as we do with the spectral bootstrapping, we then select the same number of new mock stellar masses and mock $\Delta$MSs from the sets of drawn values as are in the original grouping, allowing for repetition. We average these selected values to create a new realization of the measured quantities of our data. We repeat this 10,000 times for each grouping such that we have a large number of realizations of our data. Using the average redshift of the groupings ($z=1.6$), the newly sampled stellar masses, and newly sampled $\Delta$MS, we are able to calculate a field-like gas fraction for each new grouping of coeval field galaxies In this way, we create a sample of galaxy properties similar to our cluster galaxies and obtain mock field-like gas fractions from the T18 relation.  We Gaussian scatter the quantity log($\mu$) for each realization of ``field galaxy groupings'' assuming the inherent 0.1 dex scatter on the T18 relation\footnote{0.1 dex scatter on the T18 relation for averages of multiple galaxies, 0.2 dex scatter on the relation for individual galaxies}. If the measured gas fractions and uncertainties of our groupings are consistent with the mock ``field galaxy distributions'', we conclude that the stack is similar to field galaxies; otherwise, we conclude that the groupings of cluster members are either gas-rich or gas-deficient.  We show the forward-modeled field galaxy distributions and our measured quantities in Figure \ref{fig:FM}.

\section{Discussion}\label{sec:discussion}

\subsection{Comparison to Field and Literature} \label{subsec:self_comparison}
Looking at the groupings as a whole in Figure \ref{fig:FM}, we see that different populations of galaxies in the clusters exhibit both rich and deficient gas fractions. In this section, we examine our groupings column-by-column in Figure \ref{fig:FM}, compare the CO-detected stacks to the CO-undetected stacks, discuss the implications of scatter on values of $\alpha_{CO}$, and finally compare the results of our stacking technique to results of other intermediate-redshift and high-redshift CO and dust continuum stacking papers.\par

\subsubsection{``All'' Groupings}
Generally, we see that the ``all'' groupings (left column), which include both CO-detected and spec-z confirmed CO-undetected cluster members, are contained within the 1$\sigma$ contour of forward-modeled field-like galaxies. Despite having slight offsets from the field scaling relation (pink curves) for their weighted average stellar masses and $\Delta$MS, this indicates that, on average, cluster members do not systematically show signs of being gas-deficient or gas-rich compared to equivalent field galaxies regardless of stellar mass or $\Delta$MS. This result is notable given that the T18 field scaling relation was primarily calibrated with galaxies similar in properties to our HM-aMS grouping, and that averages of all cluster galaxies, regardless of stellar mass and $\Delta$MS, do not deviate from the relation. This indicates that on average they are consistent as a population with field galaxies.\par

\subsubsection{CO-detected Groupings}
Amongst the individually CO-detected cluster members (center column), only the HM-aMS grouping contains galaxies that are solidly consistent with field-like gas fractions. This is not wholly unexpected given the similarity in stellar masses and $\Delta$MS of our cluster members in this grouping to the field galaxies used to calibrate T18. The LM and HM-bMS groupings, however, have gas fractions $>3\sigma$ above field predictions from T18. These findings are consistent with \cite{2017ApJ...842L..21N, 2019ApJ...870...56N} which focused only on the individually CO-detected cluster members.  While it is impossible to narrow down the environmental effects on these groupings from their integrated gas mass measurements alone, these results could be indicative of enhanced molecular gas formation as a result of RPS \citep{Moretti_2020a, Moretti_2020b}, inefficient star-formation, or gas compression (see \citealt{2021PASA...38...35C} for further discussion).\par

\subsubsection{CO-undetected Groupings}
Given the field-like and gas-rich detections within the CO-detected groupings, it is unsurprising that the CO-undetected galaxies (right column) generally show gas deficiencies. The LM CO-undetected grouping (top right panel) presents an upper limit, but the limit includes a portion of field galaxy stacks at the 1$\sigma$ level. We are unable to speculate what this upper limit could mean as it only rules out the cluster members being gas-rich or consistent to within 1$\sigma$ of the field scaling relation. The two remaining HM groupings have below field-like gas fractions at $\sim2\sigma$ level.  These represent the first stacked gas mass detections for CO-undetected, high-redshift, cluster galaxies in this region of $M_{*}-\Delta$MS parameter space (see Figure \ref{fig:sample_binning}) and yet fail to find a significantly gas-deficient (more than 0.5 dex below the field scaling relation) population (-0.22 and -0.30 dex below T18, respectively). This result is significant because clusters at lower redshifts ($z<1$) have an established population of significantly gas-deficient members over the same range of $\Delta$MS that is probed at high redshift. However, 6 of the 10 galaxies in the CO-undetected HM-bMS grouping have the FIR portion of their SED constrained only by HerMES 250/350/500$\mu$m upper limits (galaxies outside of our ALMA dust continuum imaging). These galaxies may have SFRs that are higher than currently estimated which may place them in the HM-aMS grouping. Moving those galaxies out of the HM-bMS grouping could decrease the stacked CO emission, resulting in a more gas-deficient population of galaxies.  \par

\subsubsection{Comparison of CO-detected to CO-undetected in the bMS Bins}\label{subsec:co_vs_no_co}
One of the most striking results in these stacks is the dichotomy between the deficiency and enhancement of gas fractions in the HM-bMS CO-undetected and CO-detected groupings from forward-modeled field galaxies (-2$\sigma$ and +3$\sigma$).  Again, we note that the T18 field scaling relation is primarily calibrated for galaxies on or above the SFMS at this redshift. This signifies a need for gas mass measurements of field galaxies below the SFMS, especially at $z>1$. This makes interpretation challenging, but we add that most galaxies in the HM-bMS grouping have Main Sequence offsets within one dex of T18 such that we are not comparing quiescent galaxies, for which the field scaling relation does not hold, to field galaxies. \par

The stark difference between the CO-detected and CO-undetected stacks has been seen before in massive quiescent galaxies \citep{2018ApJ...860..103S, 2021Natur.597..485W, 2021ApJ...908...54W}. While we do not definitively classify our galaxies as passive/quiescent due to their observed H$\alpha$ emission and position in the UVJ diagram, our HM-bMS CO-detected and CO-undetected galaxies have very similar stacked Main Sequence offsets ($\Delta\textrm{MS}=-0.60$ and $-0.86$, respectively) to the CO-detected and CO-undetected massive quiescent galaxies from \citealt{2018ApJ...860..103S} ($z\sim0.7$, $\Delta\textrm{MS} \sim-0.7$ and $\sim-0.9$, respectively). In contrast to the gas enhancement of our HM-bMS CO-detected grouping, the CO-detected massive quiescent galaxies are consistent with T18. However, both our HM-bMS CO-undetected grouping and the CO-undetected massive quiescent galaxies are below T18 and preserve the same $\Delta$T18$\sim0.7$ dex separation from their CO-detected counterparts.  As \cite{2018ApJ...860..103S} suggests, this potentially suggests a break in the scaling relations for galaxies significantly below the SFMS, or alternatively, differing environmental effects between the two populations.\par

For example, in the HM-bMS CO-undetected grouping, the deficient molecular gas fractions could indicate gas-removal mechanisms such as strangulation or RPS in a purely gas-removal mode. The lack of gas in these galaxies could also be explained through the same gas consumption mechanisms that occur in quiescent or post-starburst field galaxies. In field studies of post-starburst galaxies, the molecular gas content of galaxies is thought to be tied to age; galaxies with ages less than 150 Myrs can still have appreciable amounts of CO, but galaxies with ages greater than 150 Myrs are remarkably devoid of CO \citep{2022ApJ...925..153B, 2022ApJ...940...39W}. We are unable to reliably measure the ages of our cluster members without spectra of the D$_{n}$4000 break.\par

Contrary to the CO-undetected grouping, the CO-detected stack is gas enhanced, which could stem from inefficient star formation (T18 is sensitive to offset from the SFMS) or efficient molecular gas formation due to gas compression or perturbed gas from RPS (e.g. \citealt{Moretti_2020b}). While literature studies have suggested that massive galaxies are more resistant to hydrodynamical environmental processes \citep{1972ApJ...176....1G, CowieSongaila1977, 1982MNRAS.198.1007N}, massive galaxies in the local Universe have exhibited gas enhancement in RPS stripped tails (jellyfish galaxies, e.g. \citealt{Moretti_2020a}). Varying stages of RPS also show connections with the SFR of the disk or tails of the galaxies \citep{2020MNRAS.495..554R} and, specifically, a connection has been observed between enhanced molecular gas in a galaxy's disk and star formation occurring in the RPS tail \citep{Moretti_2023}. Future work will focus on spatially-resolved studies of the molecular gas in our cluster galaxies to look for asymmetries associated with RPS.\par

\subsubsection{Variance of $\alpha_{\textrm{CO}}$}\label{subsec:aco}
Another pertinent consideration in these interpretations is the impact of the variance of $\alpha_{CO}$ within galaxies and amongst galaxy populations. Historically, $\alpha_{\textrm{CO}}$ has been calibrated in the Milky Way as 4.36 [(M$_{\odot}$ km s$^{-1}$ pc$^{-2}$)$^{-1}$], but has also been shown to have dependence on galaxy metallicity and stellar mass surface densities \citep{2013ARA&A..51..207B}. For all groupings in this work, we need to increase the value of  $\alpha_{\textrm{CO}}$ from the Milky Way 4.36 as a result of our stellar mass-metallicity correction (see Table \ref{tab:SED_properties} and Figure \ref{fig:stacked_spectra}). Generally, the $\alpha_{\textrm{CO}}$ correction does not change the significance in the offset from the field scaling relations and increases the $\Delta$T18 value for each grouping by 0.1 dex or less. \par

Nevertheless, we consider what values of $\alpha_{\textrm{CO}}$ would be required for our measured CO luminosities to be in line with the T18 field scaling relation. Most significantly, the HM-bMS CO-detected grouping would require an $\alpha_{\textrm{CO}}$ value of 1.74, and the HM-bMS CO-undetected grouping would need an $\alpha_{\textrm{CO}}$ value of 8.81 to be consistent with coeval field galaxies. Broadly, low values of $\alpha_{\textrm{CO}}$ are associated with starburst-type galaxies, which is not representative of the bMS galaxies, and high values of $\alpha_{\textrm{CO}}$ are associated with low mass/low metallicity galaxies \citep{2013ARA&A..51..207B}. An $\alpha_{\textrm{CO}}$ value of 8.81 for our CO-undetected HM-bMS grouping would require a 12+log(O/H) metallicity of  $\sim$8.34, which is unlikely for galaxies with log($M_{*}/M_{\odot}$) = 10.84. The disjointed characteristics of high and low values of $\alpha_{\textrm{CO}}$ and the high stellar mass with $\Delta$MS below the SFMS indicate that environmental effects are altering the gas content in these galaxies. However, without metallicity information for all cluster members or CO Spectral Line Energy Distributions, we cannot unravel the full picture of molecular gas in these cluster galaxies.\par

\subsection{Stacking Analyses Comparison} \label{subsec:stakcing_analyses_comparison}
Similar to this work, \cite{2018ApJ...856..118H} performed a stacking analysis on 12 z=1.46 star-forming cluster galaxies with spectroscopic redshifts, but without CO-detections. The constituent galaxies in this stack are distributed in a similar region on the UVJ diagram as the low mass galaxies in this work (see Figure \ref{fig:uvj}, dark green points outside the quiescent wedge). The average stellar mass of their star-forming without CO group is similar to our LM groupings (0.6$\times$10$^{10}$ $M_{\odot}$, and 0.4-0.5$\times$ 10$^{10}$ $M_{\odot}$ for our LM groupings). No detection is found for their stacked spectrum whereas we find detections for two of our three LM groupings. The difference in the non-detection from \cite{2018ApJ...856..118H} is not unexpected as the depth of the SpARCS CO(2-1) images is $\sim$10$\times$ deeper ($\sim$0.01 mJy beam$^{-1}$ vs. 0.12 mJy beam$^{-1}$ in 400 km s$^{-1}$ channels). \cite{2018ApJ...856..118H} place a 5$\sigma$ gas mass detection limit on their stacked spectrum of star-forming without CO detections at 1.05$\times$10$^{10}$ $M_{\odot}$ ($f_{gas} < 62\%$), whereas we find detections on the LM ``all'' and CO-detected groupings of 1.3 and 4.3 $\times$10$^{10}$ $M_{\odot}$ ($f_{gas} = 75\%$ and 90$\%$, respectively). The 3$\sigma$ detection limit for our LM CO-undetected grouping is 0.9$\times$10$^{10}$ $M_{\odot}$, which is indeed a lower molecular gas mass than the star-forming without CO grouping.\par

In addition to CO, dust continuum emission has also been used as a proxy for molecular gas mass. \cite{2022ApJ...927..235A} surveyed a cluster population at z$\sim$1.4 and performed an image stacking analysis on 126 cluster members. The overall dust continuum signal was found to be gas-deficient amongst galaxies with a positive offset from the star-forming main sequence. The two dust continuum stacks, binned into low mass (log($M_{*}/M_{\odot})=10.7\pm0.04)$ and high mass (log($M_{*}/M_{\odot})=11.3\pm0.05)$ groups, have main sequence offsets of 0.49 and 0.20 dex, respectively, and have offsets from the field scaling relations of -0.57 and -0.63 dex. This analysis did include cluster members with only photometric redshifts which could include galaxies that were not a part of a cluster. \cite{2022ApJ...927..235A} comment that including photometric redshift cluster members could dilute their dust continuum signal by including galaxies that were not truly at their photometric redshift. Therefore, the gas-deficient signal could be expected. The stellar masses and SFRs of the z$\sim$1.4 galaxies are similar to that of our HM-aMS groupings, for which we detect gas masses that are field-like (16 cluster members) and gas-deficient (3 cluster members). 

 Instead of binning by mass, \cite{2019ApJ...874...53B} group galaxies by their environment, with the highest environmental density classified by $>$2.6 Mpc$^{-2}$. All three stacks of dust continuum data are reported as gas-deficient with the highest density bin having the lowest gas content (-0.42 dex offset in gas fraction from T18), and is consistent with the z$\sim$1.4 stacks. This bin, despite being considered high density, is $\sim$6.5$\times$ lower density than the spectroscopic detections in the z=1.6 SpARCS clusters. This result and the results from \cite{2022ApJ...927..235A} potentially signify that dust continuum flux consistently underestimates gas mass compared to CO flux. Simulations of high mass galaxies also suggest that the dust-to-gas ratio becomes highly variable with age as dust is destroyed much faster than molecular gas due to sputtering and other processes \citep{2021ApJ...922L..30W}. \par

\cite{2022ApJ...933...11I}, in conjunction with \cite{2018ApJ...856..118H}, compared the CO-based and dust continuum-based molecular gas estimates and finding statistical agreement for 10 $z=1.46$ cluster galaxies, a robust sample of CO and dust continuum fluxes is needed to establish whether there are systematic corrections needed for the cluster environment when estimating the molecular gas content of galaxies. The work presented in this paper utilizes 21 dust continuum detections that have corresponding CO-fluxes, but we leave the comparison of the molecular gas mass estimates using CO and dust continuum for a future work (in prep.).

\section{Conclusion \label{sec:conclusion}}
We present the average molecular gas fractions of cluster galaxies in three $z=1.6$ galaxy clusters (SpARCS J0224, J0225, and J0330). We employ spectral stacking on 9 groupings of 54 cluster members with significant detections in 8 of the 9 stacks. We also employ a forward modeling analysis to extrapolate the uncertainty of our measured quantities through a molecular gas mass field scaling relation to ascertain whether our measured gas masses are similar to star-forming field galaxies. Our Low Mass and High Mass-below Main Sequence groupings probe a new parameter space of stellar masses and SFRs not covered by other high-redshift spectral stacking and dust continuum stacking studies ($\sim0.6$ dex lower in stellar mass, $\sim1$ dex lower in offset from the SFMS). To date, this is the largest CO study of cluster galaxies below the SFMS at $z>1$. Our main results are as follows:

\begin{enumerate}
    \item On average, the ensemble population of CO-detected and CO-undetected cluster galaxies have molecular gas masses consistent with field galaxies within all three of our stellar mass and $\Delta$MS groupings. We see no evidence of significant or systematic molecular gas deficiencies amongst the ensemble of high-redshift cluster members in contrast to local cluster members and stacked high-redshift dust continuum studies. The average gas fraction of galaxies with: log$(M_{*}/M_{\odot}) < 10.0$ to be 75$\pm$11\%, $log(M_{*}) > 10.0$ and $\Delta$MS $>$ -0.3 dex to be 65$\pm$8\%, and $log(M_{*}) > 10.0$ and $\Delta$MS $<$ -0.3 dex to be 32 $\pm$ 6\%.
    
    \item We recover a stacked CO(2-1) detection even when breaking down each of our three groupings further into galaxies that are CO-detected and CO-undetected, apart from one low mass grouping of CO-undetected galaxies for which we obtain an upper limit. The molecular gas fractions of these groupings range from gas-deficient ($\sim$$2\sigma$ below forward-modeled stacks of field galaxies) to gas-rich ($>3\sigma$ above forward-modeled field galaxies) perhaps indicating environmental molecular gas depletion and/or enhanced molecular gas formation. However, this is difficult to disentangle in a population analysis of integrated gas measurements.

    \item Specifically amongst high mass galaxies (log($M_{*}$) $>$ 10.0) with $\Delta$MS $<$ -0.3 dex, we find two stacks with statistically robust differences. There is significant variance from the molecular gas field scaling relation in both the gas-rich and gas-deficient directions for the CO-detected and CO-undetected stacks, respectively. This potentially signifies environmental processes most strongly affecting this population of galaxies.
    
\end{enumerate}

Many of the SpARCS cluster members also exhibit interesting CO morphologies and kinematics \citep{2019ApJ...870...56N,2023ApJ...944..213C}; as such, we save further study of this and spatially-resolved deviations from scaling relations for a future work. In addition, future studies involving the comparison of CO and dust continuum fluxes as well as the study of the molecular gas content of field galaxies below the Star-Forming Main Sequence will further elucidate the effect of environment on molecular gas.

\vspace{0.2cm}
\section{Acknowledgements}
\begin{acknowledgments}
This work was supported in part by the NRAO Student Observing Support (SOS) award SOSPADA-024. 
A.N. additionally acknowledges support from the National Science Foundation through grant AST-2307877, from HST programs GO-16300 and GO-17439, and from the Beus Center for Cosmic Foundations at Arizona State University. Support for program numbers GO-16300 and GO-17439 was provided by NASA through grants from the Space Telescope Science Institute, which is operated by the Association of Universities for Research in Astronomy, Incorporated, under NASA contract NAS5-26555. 

GW gratefully acknowledges support from the National Science Foundation through grant AST-2205189.
The work of CCW is supported by NOIRLab, which is managed by the Association of Universities for Research in Astronomy (AURA) under a cooperative agreement with the National Science Foundation.

YMB acknowledges support from UK Research and Innovation through a Future Leaders Fellowship (grant agreement MR/X035166/1).

This paper makes use of the following ALMA data: ADS/JAO.ALMA \#2017.1.01228.S, ADS/JAO.ALMA \#2018.1.00974.S, ADS/JAO.ALMA \#2021.1.01002.S, ADS/JAO.ALMA \#2021.1.01257.S. ALMA is a partnership of ESO (representing its member states), NSF (USA) and NINS (Japan), together with NRC (Canada), NSTC and ASIAA (Taiwan), and KASI (Republic of Korea), in cooperation with the Republic of Chile. The Joint ALMA Observatory is operated by ESO, AUI/NRAO and NAOJ.\par

\end{acknowledgments}




\bibliography{sample631}{}
\bibliographystyle{aasjournal}

\end{document}